\documentclass[twocolumn,aps,pra]{revtex4-1}%linenumbers
\usepackage{epsfig}
\usepackage[english]{babel}
\usepackage{latexsym}
\usepackage{subfigure}
\usepackage{graphics}
\usepackage{epstopdf}
\usepackage{dcolumn}
\usepackage{amsmath}
\usepackage{hyperref}
\usepackage{amssymb}
\usepackage{appendix}
\usepackage{color}
\usepackage{lineno}
\usepackage{soul}
\usepackage{booktabs}
\usepackage{longtable}
\usepackage{multirow}
\usepackage{array}
\usepackage{ulem}

\begin{document}
 
\title{Attosecond Reconstruction of Strain Tensors via Electronic Fingerprints}
\author{Jing Li,$^{1}$ Jiayu Yan,$^{2}$ Guoyong Yuan,$^{1}$ Chao Chen,$^{3,\ast}$ Shang Wang,$^{1,\dag}$ and Fulong Dong$^{2,\ddag}$}

\date{\today}

\begin{abstract}
We demonstrate an attosecond transient absorption spectroscopy (ATAS) scheme for reconstructing strain tensors in two-dimensional materials.
Using strained graphene as a prototype system, we show that the fishbone structures in ATAS serve as distinctive spectral fingerprints of strain, where strain-induced shifts and splittings of van Hove singularities encode the magnitude and orientation of the strain tensor, respectively.
By combining density-matrix simulations with analytical modeling, we establish a direct mapping between transient absorption spectra and strain tensors, enabling accurate retrieval of lattice deformation from ultrafast electronic responses.
Our work introduces an attosecond spectroscopic paradigm for ultrafast strain metrology, where electronic fingerprints replace conventional structural probes for sensing lattice deformation in quantum materials.

\end{abstract}
\affiliation{\textsuperscript{1}College of Physics, Hebei Key Laboratory of Photophysics Research and Application, Hebei Normal University, Shijiazhuang 050024, China}
\affiliation{\textsuperscript{2}College of Physics Science and Technology, Hebei University, Baoding 071002, China}
\affiliation{\textsuperscript{3}College of Physics and Electronic Engineering, Xingtai University, Xingtai 054001, China}

\maketitle

\textit{Introduction}. 
Strain engineering has emerged as a powerful strategy for tuning the electronic, optical, and topological properties of quantum materials by modifying their band structures \cite{GGNaumis,ZPeng,JMKim}. 
Accurate characterization of strain is therefore essential for understanding and controlling strain-induced phenomena. 
Conventional strain measurement techniques, including X-ray diffraction and Raman spectroscopy, primarily rely on structural responses, such as lattice deformation and phonon frequency shifts, to determine strain \cite{SDolabella,TKSWong}. 
However, these approaches are intrinsically limited in probing the ultrafast evolution of strain-dependent electronic structures under nonequilibrium conditions \cite{VNPopov,MAssili}. 
Developing a non-destructive approach capable of directly retrieving strain information from ultrafast electronic fingerprints remains an important challenge.

Lattice deformation modifies electronic band structures \cite{VMPereira}, suggesting that electronic fingerprints can provide an alternative route for strain metrology beyond conventional structural measurements.
Attosecond transient absorption spectroscopy (ATAS) provides a powerful approach for accessing ultrafast electronic dynamics in solids with attosecond temporal resolution \cite{MartinSchultze,MLucchini,FSchlaepfer,MVolkov,TOtobe,MatteoLucchini,GLDolso}. 
In a typical pump–probe scheme, an IR pulse drives electronic dynamics, while a time-delayed XUV pulse probes transient modifications of electronic transitions through absorption spectra \cite{TOtobe1,SYamada,PPeng,ZYin,ERidente,BRdeRoulet,LWang,JLiu,LBDrescher,YFu,DMatselyukh}. 
ATAS has been successfully applied to investigate carrier dynamics, ultrafast electronic structure evolution, and symmetry-breaking phenomena in crystal and low-dimensional materials \cite{KUchida,GioCistaro,kang2021,Dong4,GInzani,SLi,Dong5,OKneller,Dong6}. 
These capabilities open a pathway toward ultrafast strain metrology by extracting lattice information encoded in nonequilibrium electronic fingerprints.

\begin{figure}[b]
\begin{center}
{\includegraphics[width=8.5cm,height=4.5cm]{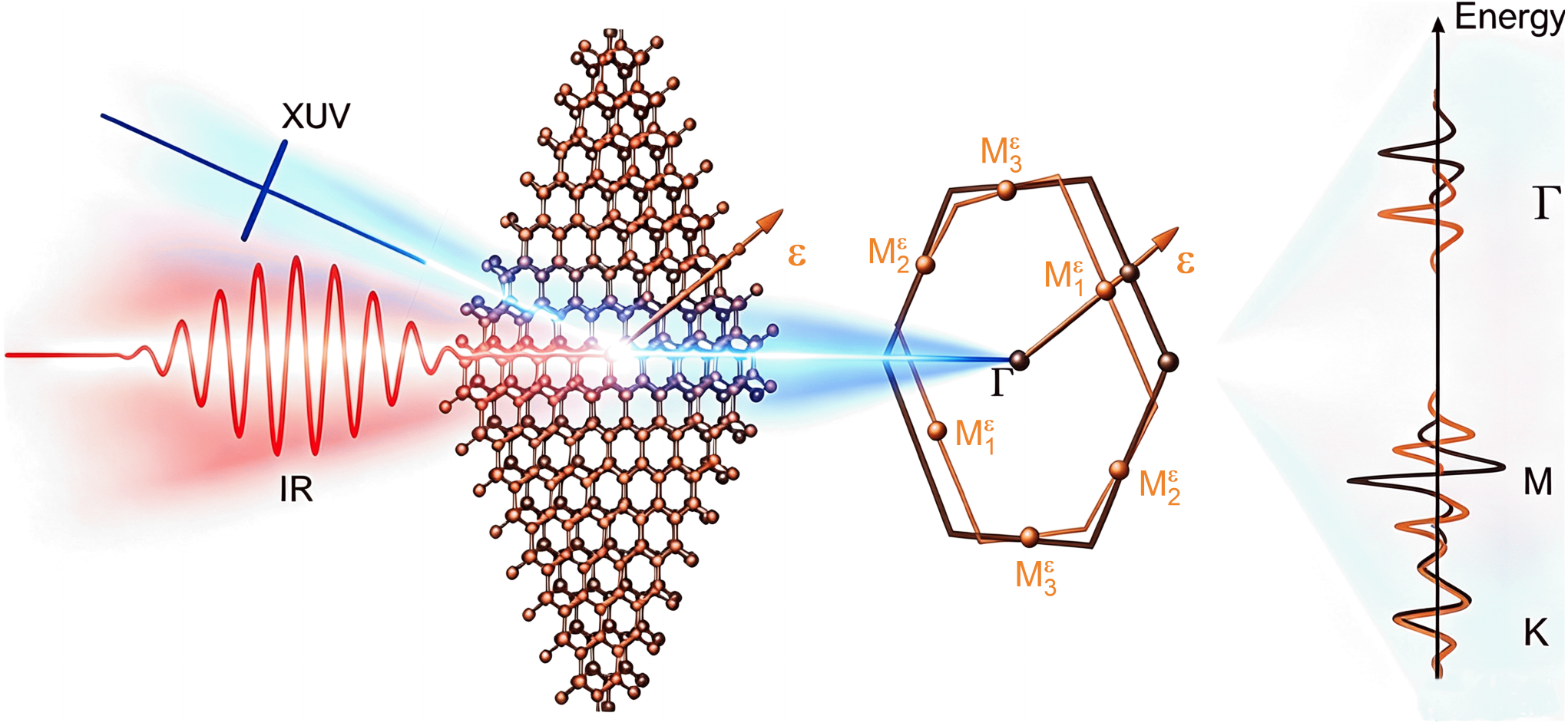}}
\caption{
Schematic illustration of ATAS for pristine and strained graphene.
An IR pump pulse and an XUV probe pulse interact with graphene under uniaxial strain.
The applied strain distorts the Brillouin zone, resulting in shifts of the van Hove singularities and corresponding changes in the transient absorption spectrum, providing electronic fingerprints for reconstructing the strain tensor.
}
\label{fig:graph1}
\end{center}
\end{figure}

Graphene offers an ideal platform for realizing such an electronic-structure-based strain metrology.
In graphene, ATAS exhibits  characteristic fishbone structures arising from strong-field-driven intraband dynamics, where the spectral features are closely associated with the van Hove singularities of the electronic band structure \cite{Dong4}. 
Meanwhile, strain engineering in graphene induces pronounced modifications of its electronic properties, including shifts of van Hove singularity energies, lifting of degeneracies, and variations in electronic band curvatures \cite{THeinrich,JZhao,KJGannan}. 
Because these strain-dependent electronic fingerprints are encoded in the transient absorption spectra, ATAS provides a potential route for retrieving lattice deformation from ultrafast electronic responses. 
In particular, strain-induced shifts and splittings of the van Hove singularities generate distinct spectral signatures, enabling the reconstruction of both the magnitude and orientation of the strain tensor.

\begin{figure*}[t]
\begin{center}
{\includegraphics[width=18cm,height=8cm]{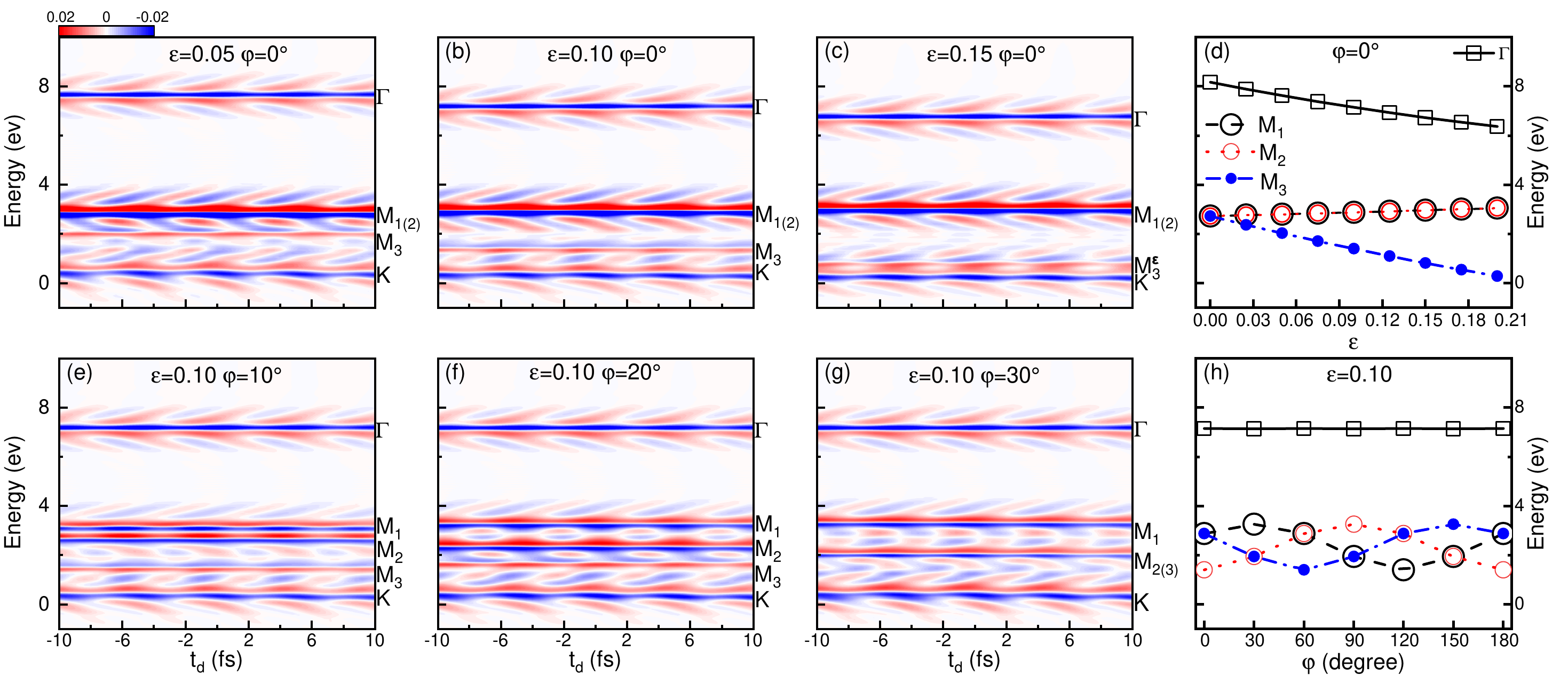}}
\caption{
(a)–(c) ATAS for strain tensors $\boldsymbol{\varepsilon} = (0.05, 0^\circ)$, $(0.10, 0^\circ)$, and $(0.15, 0^\circ)$, respectively.
(d) Energies of the $\Gamma$ and $\mathrm{M}_n^\varepsilon$ related spectral features as functions of the strain magnitude $\varepsilon$ at $\varphi = 0^\circ$.
(e)–(g) ATAS for strain tensors $\boldsymbol{\varepsilon} = (0.10, 10^\circ)$, $(0.10, 20^\circ)$, and $(0.10, 30^\circ)$, respectively.
(h) Energies of the $\Gamma$ and $\mathrm{M}_n^\varepsilon$ related spectral features as functions of the strain angle $\varphi$ at $\varepsilon = 0.10$.
The energies shown in (d) and (h) are extracted from the strain-dependent tight-binding band structure.}
\label{fig:graph1}
\end{center}
\end{figure*}

In this work, we demonstrate an ATAS-based scheme for reconstructing strain tensors in uniaxially strained graphene. 
Combining density-matrix simulations with analytical modeling \cite{SMCandLBM}, we establish a quantitative correspondence between strain tensors and ATAS spectral signatures.
We show that the fishbone structures associated with the van Hove singularities encode both the magnitude and orientation of strain, with their spectral positions and dynamical features originating from strain-induced modifications of van Hove singularity energies and electronic band curvatures. 
Based on these strain-dependent electronic fingerprints, we develop a reconstruction scheme capable of accurately retrieving the strain magnitude and orientation, establishing an attosecond spectroscopic approach for strain metrology in two-dimensional quantum materials. (Throughout the paper, atomic units are used if not specified.)

\textit{Theoretical computational models}. 
As shown in Fig. 1, graphene is subjected to a uniform uniaxial strain, which  is described by a two-dimensional strain tensor $\boldsymbol{\varepsilon}$parameterized by the strain magnitude $\varepsilon$ and orientation angle $\varphi$ between the principal strain axis and the $x$ axis .
Under strain, the dispersion relations of the conduction $(c)$ and valence $(v)$ bands are given by
$E^{\boldsymbol{\varepsilon}}_{c}(\mathbf{k}) = - E^{\boldsymbol{\varepsilon}}_{v}(\mathbf{k}) = \left| t_2 + t_3 e^{-i\mathbf{k}\cdot\mathbf{a}_1^{\boldsymbol{\varepsilon}}} + t_1 e^{-i\mathbf{k}\cdot\mathbf{a}_2^{\boldsymbol{\varepsilon}}} \right|$ \cite{VMPereira}.
where $t_i$ ($i=1,2,3$) are the strain-dependent nearest-neighbor hopping parameters, $\mathbf{a}_j^{\boldsymbol{\varepsilon}}$ ($j=1,2$) are the strained lattice basis vectors, and $\mathbf{k}$ represents the crystal momentum.
The lattice deformation modifies both the reciprocal lattice and the electronic dispersion.
Consequently, the energy at the $\Gamma$ point shifts, while the three originally equivalent $\mathrm{M}$  points become energetically inequivalent due to strain-induced symmetry breaking.
Their corresponding positions in reciprocal space are also displaced by the deformation of the Brillouin zone.
We denote these strain-induced inequivalent points as $\mathrm{M}_1^{\boldsymbol{\varepsilon}}$, $\mathrm{M}_2^{\boldsymbol{\varepsilon}}$, and $\mathrm{M}_3^{\boldsymbol{\varepsilon}}$ in Fig. 1 (see Sec. I in the Supplemental Material for details).

We simulate the ATAS of strained graphene by numerically solving the density-matrix equations in the length gauge. 
In the simulation, the pump pulse is normally incident on the graphene plane, with its electric-field polarization aligned along the  $\Gamma-\mathrm{K}$ direction, whereas the probe pulse is polarized along the out-of-plane direction.
The laser parameters are the same as those adopted in Ref. \cite{Dong4}, while the pump and probe intensities are set to  $5\times10^{10}$ W/cm$^2$ and $5\times10^{8}$ W/cm$^2$, respectively (see Sec. II in the Supplemental Material for details). 
The time delay $t_d$ is defined as the temporal separation between the maxima of the pump and probe pulse intensity envelopes.
Using the calculated time-dependent dipole, we obtain the ATAS, $\Delta S^{\boldsymbol{\varepsilon}}(\omega,t_d)$, by evaluating the deviation between the response functions with and without the pump pulse. 
For different strain tensors $\boldsymbol{\varepsilon}$, the ATAS exhibit pronounced variations, as shown in Fig. 1.
These strain-dependent spectral signatures originate from the modified electronic structure and provide electronic fingerprints for reconstructing the strain tensor.

\begin{figure*}[t]
\begin{center}
{\includegraphics[width=18cm,height=7cm]{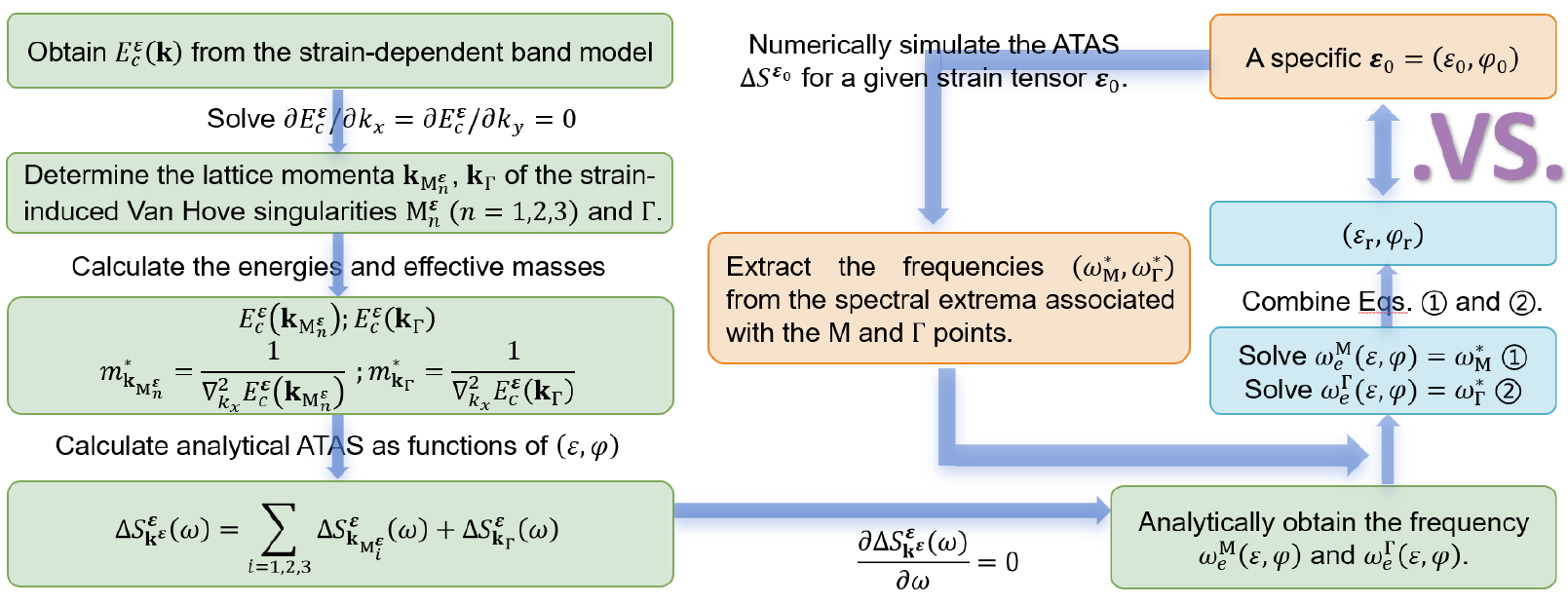}}
\caption{
Reconstruction scheme of strain parameters in uniaxially strained graphene using ATAS.
}
\label{fig:graph1}
\end{center}
\end{figure*}

\textit{Numerical results}.
In Fig. 2, we investigate the strain dependence of the ATAS $\Delta S^{\boldsymbol{\varepsilon}}(\omega,t_d)$ by fixing the strain direction at $\varphi = 0^\circ$. 
For $\boldsymbol{\varepsilon}=(0.05,0^\circ)$, four characteristic fishbone structures emerge, centered around the energies associated with the van Hove singularities at $E^{\boldsymbol{\varepsilon}}_{c}(\mathbf{k}_{\Gamma})$, $E^{\boldsymbol{\varepsilon}}_{c}(\mathbf{k}_{\mathrm{M}_1^{\boldsymbol{\varepsilon}}})=E^{\boldsymbol{\varepsilon}}_{c}(\mathbf{k}_{\mathrm{M}_2^{\boldsymbol{\varepsilon}}})$, $E^{\boldsymbol{\varepsilon}}_{c}(\mathbf{k}_{\mathrm{M}_3^{\boldsymbol{\varepsilon}}})$, and $E^{\boldsymbol{\varepsilon}}_{c}(\mathbf{k}_{\mathrm{K}})$. 
As the strain magnitude increases, the fishbone structures associated with $\Gamma$ and $\mathrm{M}_3^{\boldsymbol{\varepsilon}}$ shift toward lower energies, whereas the feature associated with the degenerate  $\mathrm{M}_{1}^{\boldsymbol{\varepsilon}}$ and $\mathrm{M}_{2}^{\boldsymbol{\varepsilon}}$ remains nearly unchanged.

To identify the origin of these strain-dependent spectral variations, we calculate the evolution of the van Hove singularity energies as a function of strain. 
As shown in Fig. 2(d), the energies at $\Gamma$ and $\mathrm{M}_3^{\boldsymbol{\varepsilon}}$ decrease monotonically with increasing tensile strain, whereas the degenerate $\mathrm{M}_1^{\boldsymbol{\varepsilon}}$ and $\mathrm{M}_2^{\boldsymbol{\varepsilon}}$ energies exhibit a slight increase. 
These trends reproduce the corresponding shifts of the fishbone structures in the ATAS, demonstrating that their spectral positions are primarily determined by the strain-dependent van Hove singularity energies.

We further explore the strain-direction dependence of the ATAS by fixing the strain magnitude at $\varepsilon=0.10$. 
A comparison between Fig. 2(e) and Fig. 2(b) reveals that a $10^\circ$ rotation of the strain direction lifts the energy degeneracy of the $\mathrm{M}_1^{\boldsymbol{\varepsilon}}$ and $\mathrm{M}_2^{\boldsymbol{\varepsilon}}$ points, resulting in the emergence of an additional spectral feature.
As the strain direction is further rotated to $20^\circ$ and $30^\circ$, the $\mathrm{M}$-point-related spectral features exhibit pronounced shifts, demonstrating that the strain orientation is encoded in the ATAS. 
We further calculate the strain-direction dependence of the van Hove singularity energies, as shown in Fig. 2(h).
The energy $E^{\boldsymbol{\varepsilon}}_{c}(\mathbf{k}_{\Gamma})$ remains nearly unchanged with varying strain direction, whereas the energies associated with the $\mathrm{M}_n^{\boldsymbol{\varepsilon}}$ points exhibit distinct periodic oscillations. 
These calculated energy variations are consistent with the shifts of the corresponding fishbone structures in the ATAS spectra, confirming that the strain orientation is encoded through the splitting and evolution of the $\mathrm{M}$-point spectral fingerprints.

The results in Fig. 2 reveal that the strain-induced evolution of the $\mathrm{M}$-point fishbone structures in ATAS encode both the magnitude and orientation of the applied strain. 
Their spectral positions exhibit pronounced variations with the strain tensor, reflecting the anisotropic modification of the electronic structure around the $\mathrm{M}$ points.
In contrast, the $\Gamma$-point spectral feature depends only on the strain magnitude and remains insensitive to the strain direction, which originates from the isotropic nature of the band dispersion around the $\Gamma$ point. 
These distinct strain-dependent spectral responses provide the essential information for reconstructing the strain tensor from the ATAS.
In the following, we reconstruct the strain tensor parameters using the spectral energies associated with the $\mathrm{M}$ and $\Gamma$ points. 
(A complementary reconstruction scheme based solely on the spectral energy and intensity of the $\mathrm{M}$ point is presented in Sec. IV of the Supplemental Material.)

\begin{figure}[t]
\begin{center}
{\includegraphics[width=9cm,height=16cm]{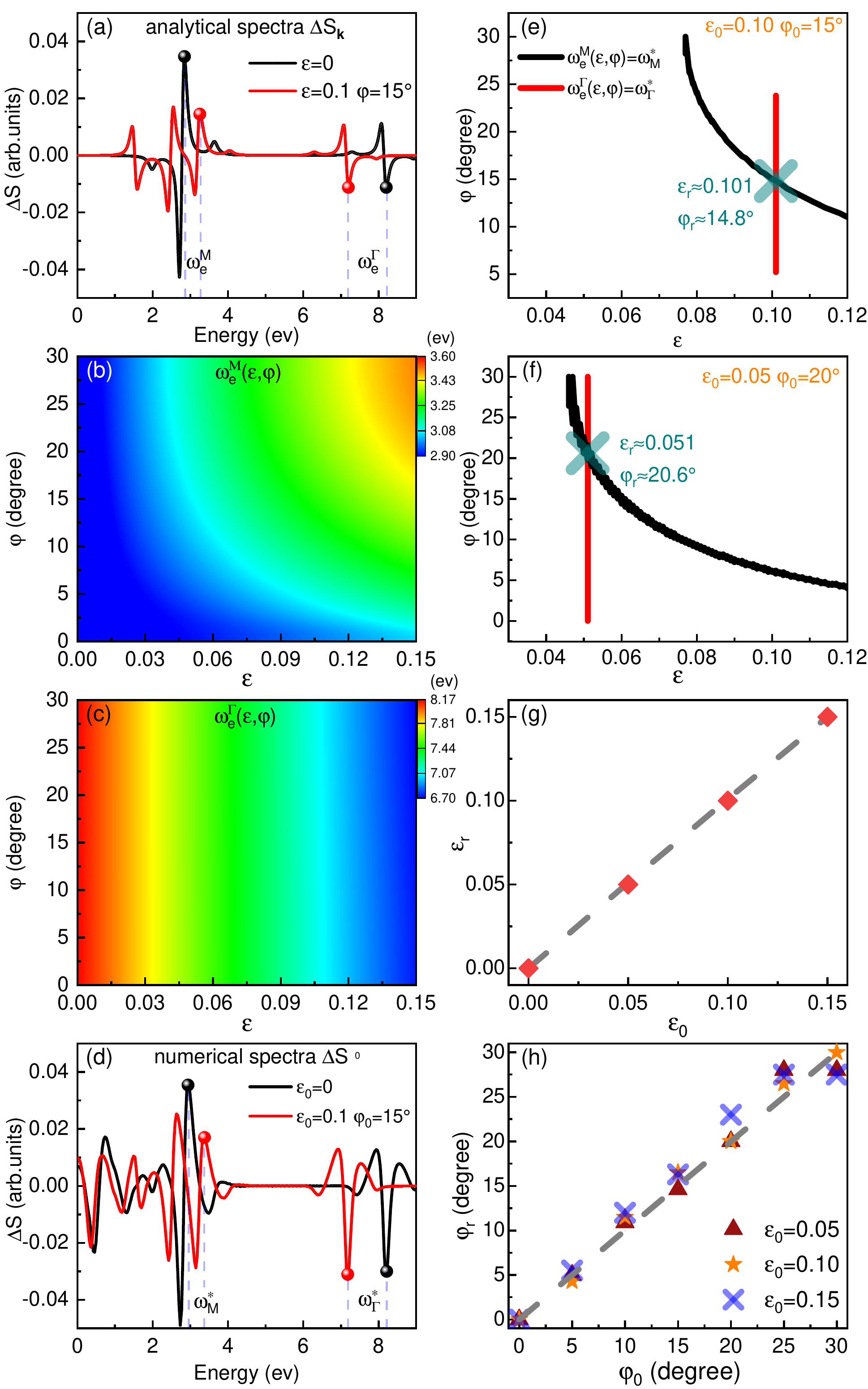}}
\caption{
(a) Analytical ATAS spectra of pristine and strained graphene. 
The spheres indicate the characteristic frequencies $\omega_e^\mathrm{M}$ and $\omega_e^\Gamma$.
(b) and (c) Mappings of the characteristic frequencies as functions of the strain parameters. 
(d) Numerically simulated ATAS for pristine and strained graphene.
Corresponging characteristic frequencies $\omega_\mathrm{M}^{*}$ and $\omega_\Gamma^{*}$ are also marked. 
(e) and (f) Reconstruction of two representative strain states, $\varepsilon_0=(0.1,15^\circ)$ and $\varepsilon_0=(0.05,20^\circ)$. 
(g) and (h) Reconstruction deviations of the strain magnitude and orientation, respectively.
}
\label{fig:graph1}
\end{center}
\end{figure}

\textit{Reconstruction scheme and result}. 
Figure 3 illustrates our scheme for reconstructing the strain tensor from the ATAS.
For a given strain configuration $\boldsymbol{\varepsilon}$, the dispersion $E_c^{\boldsymbol{\varepsilon}}(\mathbf{k})$ is first obtained from the strained-dependent tight-binding model.
The lattice momenta of the strain-induced van Hove singularities, $\mathbf{k}_{\mathrm{M}_n^{\boldsymbol{\varepsilon}}}$, $\mathbf{k}_{\Gamma}$ are then determined by solving the conditions $\partial E^{\boldsymbol{\varepsilon}}_{c}/ \partial k_{x} = \partial E^{\boldsymbol{\varepsilon}}_{c} / \partial k_{y}=0$.
The corresponding  energies and electron effective masses projected along the IR-field polarization direction (i.e., $x$ direction $ m_{\mathbf{k}}^*=\frac{1}{\nabla_{\mathrm{k}_{x}}^2 E_c^{\boldsymbol{\varepsilon}}(\mathbf{k})}$),  can be evaluated from the strain-dependent band model.
The strain-dependent ATAS spectrum can be obtained by summing the contributions from these van Hove singularity points, $\Delta S_{\mathbf{k}^{\boldsymbol{\varepsilon}}}^{\boldsymbol{\varepsilon}}(\omega)=\sum_{i=1}^{3}\Delta S_{\mathbf{k}_{\mathrm{M}^{\boldsymbol{\varepsilon}}_i}}^{\boldsymbol{\varepsilon}}(\omega)+\Delta S_{\mathbf{k}_{\Gamma}}^{\boldsymbol{\varepsilon}}(\omega)$, where for $\Gamma(\mathrm{M})$ point, $\Delta S_{\mathbf{k}_{\Gamma (\mathrm{M})}}^{\boldsymbol{\varepsilon}}(\omega) $ represents the analytical spectrum at the delay of $t_d=0$. 
(The detailed derivation are provided in Sec. III of the Supplemental Material.)
For two representative strain configurations, $\boldsymbol{\varepsilon}=0$ and $\boldsymbol{\varepsilon}=(0.1,15^\circ)$, the corresponding analytical spectra $\Delta S_{\mathbf{k}^{\boldsymbol{\varepsilon}}}^{\boldsymbol{\varepsilon}}(\omega)$ are shown in Fig. 4(a).

We extracted the characteristic frequencies by identifying the highlighted extrema satisfying $\partial \Delta S_{\mathbf{k}^{\boldsymbol{\varepsilon}}}^{\boldsymbol{\varepsilon}}(\omega)/\partial\omega=0$.
Among the extrema associated with the $\mathrm{M}_n^{\boldsymbol{\varepsilon}}$ points, the one with the highest energy is selected. 
This criterion minimizes the contribution from absorption spectra near the K points, ensuring that the extracted frequencies are primarily determined by electronic contributions from the $\mathrm{M}_n^{\boldsymbol{\varepsilon}}$ points. 
For the $\Gamma$ point, the most negative spectral extremum is selected. 
The corresponding characteristic frequencies are defined as $\omega_e^{\mathrm{M}}$ and $\omega_e^{\Gamma}$, marked by the spheres in Fig. 4(a). 
The resulting mappings, $\omega_e^{\mathrm{M}}(\varepsilon,\varphi)$ and $\omega_e^{\Gamma}(\varepsilon,\varphi)$, shown in Figs. 4(b) and 4(c), respectively, establish the relationship between the ATAS spectral fingerprints and the strain parameters, enabling the reconstruction of the strain parameters from the simulated spectra.

On the other hand, for a given strain tensor $\boldsymbol{\varepsilon}_0=(\varepsilon_0,\varphi_0)$, we numerically calculate the ATAS spectrum $\Delta S^{\boldsymbol{\varepsilon}_0}(\omega)$ by solving the density-matrix equations at the condition of  $t_d = 0$ used in Fig. 2. 
Similar to analytical results, the characteristic frequencies $\omega_{\textrm{M}}^*$ and $\omega_\Gamma^*$ are extracted from the extrema of the numerical spectra, as indicated by the spheres in Fig. 4(d).

By solving the equation $\omega_e^{\mathrm{M}}(\varepsilon,\varphi) = \omega_{\mathrm{M}}^*$, one can obtain the first constraint relation between the strain magnitude $\varepsilon$ and direction $\varphi$. 
Similarly, the second constraint relation is obtained from $\omega_e^{\Gamma}(\varepsilon,\varphi)=
\omega_\Gamma^*$. 
Their intersection in the $(\varepsilon,\varphi)$  parameter space determines the reconstructed strain parameters $(\varepsilon_r,\varphi_r)$. 
In Figs. 4(e) and 4(f), the black and red curves represent the two constraint relations, while the cyan crosses denote the reconstructed strain configurations . 
The agreement between the reconstructed parameters  $(\varepsilon_r,\varphi_r)$ and the original values  $(\varepsilon_0,\varphi_0)$ used in the numerical simulations demonstrates the feasibility of the proposed reconstruction scheme.

Finally, we evaluate the accuracy of the proposed reconstruction scheme. 
As demonstrated above, the strain magnitude can be accurately extracted from strain-induced shift of the $\Gamma$-point spectral feature, which exhibits negligible dependence on the strain orientation.
Figure 4(g) shows the reconstructed strain magnitudes for a series of applied strain configurations.
The gray dashed line indicates the ideal reconstruction, and the deviation remains within $\Delta\varepsilon = |\varepsilon_r-\varepsilon_0| <0.001$.
Figure 4(h) presents the reconstructed strain orientations for three representative strain magnitudes, $\varepsilon = 0.05$, $0.10$, and $0.15$. 
The deviations between the reconstructed and input strain orientations are within $\Delta\varphi= |\varphi_r-\varphi_0|<3^\circ$. 
These results demonstrate the high accuracy of the ATAS-based reconstruction scheme, corresponding to picometer-scale sensitivity to lattice distortions in two-dimensional materials.

\textit{Summary.}
We establish an attosecond electronic metrology approach for reconstructing strain parameters in two-dimensional quantum materials. 
Using uniaxially strained graphene as a prototype system, we demonstrate that ATAS fishbone structures provide distinctive electronic fingerprints of strain through strain-induced modifications of van Hove singularities and electronic effective masses. 
An analytical model establishes the microscopic connection between strain parameters and ATAS spectral signatures, enabling quantitative inversion from ATAS spectra to strain parameters. 
By combining analytical mappings with density-matrix simulations, we reconstruct strain parameters with high accuracy from ATAS spectra. 
These results demonstrate that ATAS can be transformed from a probe of ultrafast electronic dynamics into an all-optical metrology tool with picometer-scale sensitivity to lattice distortions in two-dimensional materials.

\section*{ACKNOWLEDGMENTS}

This work is supported by the National Natural Science Foundation of China (Grant No. 12404394 and No. 12347165), the Natural Science Foundation of Hebei Province, China (Grant No. A2026201012), the Science and Technology Project of Hebei Education Department (Grant No. QN2026281), the Hebei University Natural Science Interdisciplinary Research Project (DXK202510), and the High-Performance Computing Center of Hebei University.

\end{document}